\documentstyle[12pt]{article}

\def\thefootnote{\fnsymbol{footnote}}

\newcommand{\eq}{\begin{equation}}
\newcommand{\en}{\end{equation}}
\newcommand{\eqa}{\begin{eqnarray}}
\newcommand{\ena}{\end{eqnarray}}

\begin{document}
\begin{titlepage}
\begin{flushright}
HUB-EP-98/41  \\
\end{flushright}
\begin{center}
{\Large\bf Speeding up finite step-size updating of full QCD
	   on the lattice}
\end{center}
\vskip 0.8cm
\centerline{
 M. Hasenbusch\footnote{e--mail: hasenbus@physik.hu-berlin.de}}
 \vskip .2 cm
 \centerline{\sl  Humboldt Universit\"at zu Berlin, Institut f\"ur Physik}
 \centerline{\sl Invalidenstr. 110, D-10115 Berlin, Germany}
 \vskip 0.6cm

\begin{abstract}
 We propose various improvements of finite step-size updating 
 for full QCD on the lattice that might turn finite step-size updating 
 into a viable  alternative to the hybrid Monte Carlo algorithm.
 These improvements are noise reduction of 
 the noisy estimator of the fermion determinant,  unbiased inclusion of 
 the hopping parameter expansion and a multi-level Metropolis scheme.
 First numerical tests are performed for the 2 dimensional
 Schwinger model with two flavours of Wilson fermions and for QCD
 with two flavours of Wilson fermions and Schr\"odinger functional 
 boundary conditions.
\vskip0.1cm
\end{abstract}
\end{titlepage}

\setcounter{footnote}{0}
\def\thefootnote{\arabic{footnote}}

\section{Introduction}
The incorporation of  fermionic degrees of freedom
in the simulation of lattice
QCD is a longstanding problem. 
At present the  hybrid Monte Carlo algorithm \cite{DuKePeRo} is  the 
state of the art algorithm for the simulation of full QCD on the lattice.
Most of QCD simulations up to now were performed in the so called
quenched approximation where the fermion determinant 
is approximated by a constant factor.
Simulations of 2-flavour QCD on lattices of a size that might just allow a
physically meaningful interpretation were performed just   
recently \cite{japan,sesam,edinbourgh,maw}.

Extremely large autocorrelation times of the topological charge
have been observed in hybrid Monte Carlo simulations of QCD  with 
staggered fermions \cite{AlBoDEDiVi}. One might ask whether the small 
step-size of the hybrid Monte Carlo creates particular problems 
in switching 
the topological sector. The authors of ref. \cite{viele} found 
that the topological charge is indeed the slowest mode in the 
hybrid Monte Carlo simulation of QCD with Wilson fermions. However the
integrated autocorrelation time of the topological charge is only 
larger by a small factor than that of other quantities.
Nevertheless it seems desirable to have a finite step-size  
updating algorithm as a complement of the hybrid Monte Carlo.

Formally the multi-boson approach of M.  L\"uscher \cite{Lu} allows 
for a finite-step-size updating of the gauge-field. However in the 
chiral limit the number of bosonic fields has to be increased. 
These fields amount to a large "force" on the gauge-field
and allow only small changes in a single update step. See refs. 
\cite{BuJaJeLuSiSo,AlBoFoGaJe,BoFoGa}.

In fact the first proposal for a practical QCD algorithm 
by Weingarten and Petcher \cite{WePe} in 1981 is
a finite step-size algorithm. Since the update 
of a single link (or a fixed small number of links) requires
the evaluation of 
the inverse of the fermion matrix applied to a vector, the CPU-time
required for a full sweep over the lattice increases with
$Volume^2$ even at fixed $\beta$ and $\kappa$. 

For that reason this algorithm and variants of it 
\cite{FuMaPaRe,KeKu,KeKuMePe,HoKeMePe} (and many more refs.)
where abandoned when  the
hybrid Monte Carlo algorithm \cite{DuKePeRo}
was introduced in 1989 which has a volume dependence 
like $Volume^{5/4}$ for fixed $\beta$ and $\kappa$.

In this paper we will demonstrate how finite step-size algorithms 
can be speeded up by a large (order 100) factor. Still progress 
is needed to overcome the $Volume^2$ increase of the CPU-time,
such that the algorithm
becomes an alternative to the hybrid Monte Carlo algorithm
in present day simulations.

The paper is organised as follows.
In section 2 we will discuss the action to be simulated. 
Here we shall explain how the hopping parameter expansion 
can be incorporated
into the simulation in an unbiased form.
Next we show how the variance of the noisy estimator of the 
fermion determinant can be reduced. 
The major novelty of the simulation is the use of a sequence of 
approximations 
of the exact action in a multi-level Metropolis scheme (section 3).
In section 4 we will present first numerical  tests
of the methods proposed. 
These tests are performed with the two flavour 2D Schwinger model with 
Wilson action and two-flavour
QCD with Wilson action and Schr\"odinger functional boundary conditions.
In section 5 we compare our method with related approaches. Finally
we give a short outlook on possible improvements of the methods 
discussed.

\section{The action to be simulated}
In order to perform numerical simulations the Grassmann variables 
in the path-integral formulation of QCD 
are in general integrated out. What remains 
is a Boltzmann-factor that only depends on the gauge degrees of freedom.
For example in the case of two flavours of mass-degenerate fermions
we obtain
\begin{equation}
Z = \int \mbox{D} [U] \; \exp(-S_G[U]) \; {\mbox{det}} \;
M^{\dag} M \;\; ,
\end{equation}
where $S_G[U]$ is the gauge action and $ M$ the fermion matrix in its
non-hermitian form.

For reasonably large lattice sizes the problem still remains intractable
in this form since the evaluation of the determinant requires 
of the order $Volume^3$ operations.

A way out of the problem was proposed by Weingarten and Petcher who
use the identity
\begin{equation}
  \mbox{det} \;  M^{\dag} M \propto
  \int \mbox{D} [\eta] \mbox{D} [\eta^{\dag}]  \exp(-|M^{-1} \eta|^2)
\end{equation}
to introduce auxiliary bosonic degrees of freedom:
\begin{equation}
Z = \int \mbox{D} [U]  \mbox{D} [\eta] \mbox{D} [\eta^{\dag}] \;
\exp(-S_G[U]-|M^{-1} \eta|^2) \; \;. 
\end{equation}
This reduces the calculation of the action to a problem which takes of 
the order
$Volume$ operations at the price of a noisy 
estimate of the fermion matrix. 
Note that the hybrid Monte Carlo is also based on this action.

In the following we will use the hopping parameter expansion to 
evaluate part of the fermion determinant non-noisy while the remainder
is treated in a similar fashion as Weingarten and Petcher treat the full
fermion matrix.

\subsection{Making use of the hopping parameter expansion}

Let us start the discussion with the effective weight for two flavours of
degenerate fermions
\begin{equation}
\mbox{det} \;  M^{\dag} M \;\;=\mbox{det} \; M^{\dag} \;\;
\mbox{det} \; M 
= \exp({\mbox{tr}} \; \ln M^{\dag} + \mbox{tr} \; \ln M) \;\; .
\end{equation}
$M$ can be written as
\begin{equation}
 M = 1 - \kappa H
\end{equation}
(in the red-black preconditioned case we have to replace 
$\kappa$ by $\kappa^2$).
$\ln M$  can now be written as a Taylor series in $\kappa$
\begin{equation}
\ln M = -  \kappa H - \frac{1}{2} \kappa^2 H^2 - \frac{1}{3} 
\kappa^3 H^3 \;\;\;  ...  \;\; .
\end{equation}
In the following we use the first $k$ terms explicitly as action
for our updates of the link variables
while the remaining  part is dealt with stochastically.

We define
\begin{equation}
\ln \tilde M  = \ln M + \sum_{n=1}^k \frac{1}{n} \kappa^n H^n
\end{equation} 
or equivalently
\begin{equation}
 \tilde M  =  M \exp\left( \sum_{n=1}^k \frac{1}{n} \kappa^n H^n\right) \;\; .
\end{equation}
So we arrive at the action 
\begin{equation}
S[U,\eta] =
 S_G[U] + \mbox{tr} \sum_{n=1}^k \frac{1}{n} \kappa^n H[U]^n  + 
 \mbox{h.c.}
 + |\tilde M[U]^{-1} \eta|^2  \;\; .
\end{equation}
In the following section we will give a noise reduced  replacement of
$ |\tilde M^{-1} \eta|^2$
which will lead to the action which is simulated at the end.

\subsection{Reducing noise by using roots of the fermion matrix}
The fermionic determinant can be rewritten as
\begin{equation}
 \mbox{det}  M^{\dag} M =  \mbox{det}  M^{\dag}  \mbox{det} M 
 = (\mbox{det}  M^{1/r \dag} )^r (\mbox{det}  M^{1/r} )^r
 =(\mbox{det} M^{1/r \dag} M^{1/r} )^r
\end{equation}
which allows to rewrite each of the factors separately as an integral
over auxiliary bosonic variables
\begin{equation}
 \label{noise}
 (\mbox{det} M^{1/r^{\dag}} M^{1/r}  )^{r} \;\;   \propto
 \int {\mbox D} [\eta_1] {\mbox D} [ \eta_1^{\dag}] ...  
  {\mbox D} [\eta_r] {\mbox D} [\eta_r^{\dag}]
 \exp(-\sum_{i=1}^{r} | M^{-1/r} \eta_i |^2) \;\; .
\end{equation}
The $M^{1/r}$ and $M^{-1/r}$ are computed as Taylor series in $\kappa H$.

In the limit $r \rightarrow \infty $ the 
integrand gives, up to a factor that does not depend on the
gauge-field, the fermion determinant.

Note that for complex $x$ in a unit circle around  $1$ we have
\begin{equation}
\lim_{r \rightarrow \infty} \;\; r \; \left((1-x)^{1/r} -1 \right) = 
-\lim_{r \rightarrow \infty} \;\; r \; \left((1-x)^{-1/r} -1 \right)=
\ln(1-x) \;\; .
\end{equation}
This can be easily shown:
With $\exp(\tilde x) := 1-x$ we get
\begin{equation}
  r \;\; \left((1-x)^{1/r} -1 \right) = 
  r \;\; (\exp(\tilde x/r) -1) = \tilde x + O(1/r)
\end{equation}

Equilibrium $\eta_i$ for a given gauge field  are given by
\begin{equation}
 \eta_i = M'^{1/r} \chi_i = \chi_i + \frac{1}{r} \ln M' \; \chi_i +O(1/r^2)
\end{equation}
where the $\chi_i$ have a Gaussian distribution.

Hence we obtain for the integrand of eq. \ref{noise}
\begin{eqnarray}
& & \exp\left(-\sum_{i=1}^{r} | M^{-1/r} \eta_i |^2 \right) 
\nonumber \\
&=&\exp\left(-\sum_{i=1}^{r} | (1-\frac{1}{r} \ln M) \eta_i
+O(1/r^2) |^2 \right)
\nonumber \\
&=&\exp\left(-\sum_{i=1}^{r} | (1-\frac{1}{r} \ln M) \chi_i + 
\frac{1}{r} \ln M' \; \chi_i  +O(1/r^2) |^2 \right)
\nonumber \\
\end{eqnarray}
With
\begin{equation}
\lim_{r \rightarrow \infty } \frac{1}{r} \; \sum_{i=1}^{r} \;\;
\chi^{\dag}_i \; \ln M \; \chi_i = \mbox{tr} \ln M \;\;,
\end{equation}
we arrive at the result
\begin{equation}
\lim_{r \rightarrow \infty}
\exp\left(-\sum_{i=1}^{r} | M^{-1/r} \eta_i |^2 \right) 
= const \times \mbox{det} M^{\dag} M \;\; .
\end{equation}

Finally we arrived at 
the action (up to red-black preconditioning)
which is used for the simulations:
\begin{equation}
S =  S_G 
  + \sum_{n=1}^k \frac{1}{n} \kappa^n \mbox{tr} \left( H^n + h.c.\right)
  + \sum_{i=1}^{r} | \tilde M^{-1/r} \eta_i |^2 \;\; .
\end{equation}
The parameters that characterise the action 
are the order of the hopping parameter expansion $k$ and
the root $r$ of the modified fermion matrix $\tilde M$.

\section{Hierarchy  of acceptance steps}
The novel feature of our updating scheme is that a sequence of 
approximations 
\begin{equation}
S_1, S_2 , ... , S_l=S
\end{equation}
of the full action is used. This sequence of actions 
is organised such that the actions become better approximations of the 
full action while at the same time the computational effort to compute 
them increases. 

For the action above the sequence of approximations is realized in a 
rather trivial way:

The first approximation is given by the  gauge action plus the "hopping
part" of the fermion action
\begin{equation}
S_1 = S_G + 
\sum_{n=1}^k \frac{1}{n} \kappa^n \mbox{tr} \left( H^n + h.c.\right)
\;\; ,
\end{equation}
while better approximations are given by the truncation of 
the Taylor series of $ \tilde M^{-1/r} \eta_i $ at a finite order $t$.

Hence the sequence of actions is characterised by a sequence of truncation
orders  
\begin{equation}
 t_2 < t_3 <  ... < t_l ="\infty"
\end{equation}
where $"\infty"$ means that the series is truncated at an order such that 
the sum of the remaining terms is below a given (small) bound.

We should note that the auxiliary field $\eta$ can be updated
in a global heat-bath step
\begin{equation}
 \eta_j = \tilde M^{1/r} \chi_j
 \end{equation}
where $\chi_j$ has a Gaussian distribution. Therefore we will update
the  $\eta_j$ just once for every update cycle. In the following 
discussion of the update of the gauge-field $U$ we therefore assume 
a fixed $\eta_j$.

\subsection{The simplest case: $l=2$}
Let us first discuss the algorithm for the simplest case which is given by 
$l=2$.  Such two-step  decompositions of the action
are actually in common use.

For the simplicity of the discussion we
assume $r=1$ and $k=0$.
First 
a number of link-updates are performed with Cabibbo-Marinari \cite{cama}
updating 
or micro-canonical over-relaxation \cite{adler,over}
such that the whole update
sequence 
respects detailed balance with respect to the pure gauge action.
(For a detailed
discussion see below.)
This way a proposal $U'$ for the full action is  generated.
The proposal $U'$ is accepted with the probability
given by
\begin{equation}
A(U',U) =\mbox{min}[1,\exp(-s_2[U'] + s_2[U] ) ]
\end{equation}
with $s_2=S_2-S_1$ and
\begin{equation}
-s_2[U']+s_2[U]
 =
 -|M(U')^{-1} \eta|^2  + |M(U)^{-1} \eta|^2
\end{equation}

It is easy to see that this algorithm fulfils detailed balance with 
respect to the action $S_2$:

For $U \ne U'$ we get:

{\bf Case 1:}  $s_2[U'] \ge s_2[U]$ \\
For the update of $U$ to $U'$ we obtain 
\begin{equation}
P_2(U',U) = P_1(U',U) \; \exp(-s_2[U']+s_2[U])
\end{equation}
while for the update of $U'$ to $U$ we get
\begin{equation}
P_2(U,U') = P_1(U,U')  \;\; . 
\end{equation}
Taking the ratio and using the fact that $P_1$ satisfies detailed
balance with 
respect to $S_1$ we get
\begin{equation}
\frac{P_2(U',U)}{P_2(U,U')}
=\exp(-S_1[U']+S_1[U]) \; \exp(-s_2[U']+s_2[U]) = \exp(-S_2[U']+S_2[U])
\end{equation}

{\bf Case 2:} $s_2[U'] < s_2[U]$ works just analogously.

\subsection{Generalisation to $l>2$}
The generalisation to $l>2$ is done recursively. Given an update algorithm 
$UP_{i-1}$ that
fulfils detailed balance with respect to an action $S_{i-1}$
we construct an algorithm $UP_{i}$ that fulfils detailed 
balance with respect to the action $S_{i}$.
This process is iterated until 
we reach the exact action of the model.

We use the algorithm $UP_{i-1}$ that fulfils detailed balance with 
respect to the action $S_{i-1}$ in order to construct a proposal.
Starting from a configuration $U$
we apply the elementary update step of $UP_{i-1}$
$m_{i-1}$ times to obtain the proposal $U'$. The composition of the 
elementary update steps has to be done in such a way that detailed 
balance is maintained for the whole sequence. (A discussion of 
this point is given below.)
The proposal $U'$ is then accepted with the probability
\begin{equation}
A_i(U',U) =\mbox{min}[1,\exp(-s_i[U'] + s_i[U] ) ] \;\;
\end{equation}
with $s_i = S_i -S_{i-1}$.

\subsection{Composing updates}
Next we have to discuss how the updates with the action $S_1$ should
look in detail.  The basic building blocks in the case of QCD 
will be the well known 
Cabibbo-Marinari heat-bath update \cite{cama} and the micro-canonical 
over-relaxation update \cite{adler,over}.
Both algorithms fulfil detailed balance
when applied to a single subgroup of a given link-variable.

However one should note that a sequence of (different) updating steps
which individually 
satisfy detailed balance, in general does not satisfies detailed balance 
as a whole.
This statement in particular applies to sweeping through
the lattice with Cabibbo-Marinari or over-relaxation in a given order.

In simulations of the pure gauge theory this does not pose a problem since 
the sequence still satisfies the weaker and sufficient condition of
stability
\begin{equation}
 \exp(-S(U')) = \int \mbox{D} U  \;\; P(U',U) \; \exp(-S(U)) \;\; . 
\end{equation}
However the basic building-blocks of our algorithm have to satisfy
detailed balance. 
The simplest composition that does fulfil detailed balance is to select
the link and the sub-group  randomly  for each link-update.

For performance reasons however it is desirable to stay with 
a regular pattern of sweeping through the lattice.
As simple way to achieve this is to choose with probability $1/2$ 
either a given sequence of elementary update steps or with equal 
probability its exact reverse (see for example ref.  \cite{BoFoGa})
This symmetrisation restores detailed
balance.
Let as prove the statement for a sequence of two updates.

The update probability of the symmetrised composite of the two updates 
$p_1$ and $p_2$ is given by
\begin{equation}
 P(U'',U) = \frac12 \int dU' \left (p_1(U'',U') \; p_2(U',U)
	  +  p_2(U'',U') \; p_1(U',U) \right)
\end{equation}

Now for any intermediate configuration $U'$ we have:
\begin{eqnarray}
& & \frac{p_1(U'',U') \; p_2(U',U)+p_2(U'',U') \; p_1(U',U)}
      {p_1(U,U') \; p_2(U',U'')+p_2(U,U') \; p_1(U',U'')} \nonumber \\
 &=&    \exp(-S(U'')+S(U')) \; \exp(-S(U')+S(U) )   \nonumber \\
 &=&      \exp(-S(U'')+S(U)) \;\; .
\end{eqnarray}
where we have used that $p_1$ and $p_2$ satisfy detailed balance.
Since that ratio is identical for any intermediate configuration 
$U'$, the ratio of the integral over the $U'$ takes the same value as 
for each of the individual $U'$.
Hence  detailed balance is satisfied for the whole sequence.

\section{Numerical results}
As first tests of the algorithm proposed above we simulated the 
1+1-dimensional 2-flavour Schwinger model with Wilson fermions and 
3+1-dimensional 2 flavour QCD with Wilson fermions 
and with Schr\"odinger-functional boundary conditions.  In both cases 
we performed most of the simulations at one set of parameters. 
The sets of parameters where chosen such that we could compare our 
results with the literature \cite{IrSe} in the case of the Schwinger 
model are with results obtained within the Alpha-collaboration
\cite{uprivate,kprivate} in the case of QCD. 
The aim of this numerical study is to
obtain a first impression of the effectiveness of our new proposals 
compared to the hybrid Monte Carlo and the multi-boson algorithm.
Since it is very likely 
that further substantial improvements of the algorithm can be found 
we think that it is not the time yet to systematically study 
the dependence of the performance of the algorithm on all parameters
of the theory.

\subsection{The Schwinger model in 2 dimensions}
We simulate the two-flavour two dimensional lattice Schwinger model with
Wilson fermions.
The gauge part of the action is given by
\begin{equation}
   S_G= -\beta \sum_{x} \mbox{Re} \; U_{plaq,x}\;\;
\end{equation}
where 
\begin{equation}
U_{plaq,x} = U_{x,1} \;\; U_{x+(1,0),2} \;\; \bar{U}_{x+(0,1),1}
 \;\; \bar{U}_{x,2} \;\;\; ,
\end{equation}
where the link variables $U_{x,\mu}$ are elements of $U(1)$. 
The fermion matrix can be written as
\begin{equation}
 M=1 - \kappa H \;\;.
\end{equation}
The hopping part of the fermion matrix is given by
\begin{equation}
H= \sum_{\mu} \left(\delta_{x-\hat \mu,y} \; (1+\gamma_{\mu}) \;
  U_{x-\hat \mu,\mu}
    +\delta_{x+\hat \mu,y} \; (1-\gamma_{\mu}) \; \bar{U}_{x,\mu}
    \right)
\end{equation}
where in two dimensions we choose the  $\gamma$-matrices as
\begin{equation}
\gamma_1 =   \left( \begin{array}{cc}
             0   &  1  \\
             1   &  0
\end{array} \right)
\;\;\;\;\;\;\;\;\;
\gamma_2 =   \left( \begin{array}{cc}
           1   &  0  \\
	   0   & -1
\end{array} \right)
\;\; .
\end{equation}
As for other algorithms preconditioning improves the performance 
of our algorithm considerably. A simple version of preconditioning 
is the so called 
red-black preconditioning. 
The sites of the lattice are decomposed in even and odd sites. Then
the fermion matrix can be written in the from 
\begin{equation}
 M=  \left( \begin{array}{cc}
	      1_{ee}   &  -\kappa H_{eo}  \\
              -\kappa H_{oe}   &  1_{oo}
	   \end{array} \right)
\end{equation}
For the fermion determinant the identity
\begin{equation}
 \mbox{det} M = \mbox{det} (1_{ee}-\kappa^2 H_{eo} H_{oe})
\end{equation}
holds.
Hence the original problem is reduced by half in 
the dimension (of the fermion determinant).  
The bosonic field $\eta$ which is used for the stochastic estimate 
of the fermion determinant lives 
only on even sites. The red-black preconditioned fermion matrix is 
given by
\begin{equation}
M_{ee} = 1_{ee} - \kappa^2 \;\; H_{eo} \; H_{oe} \;\; .
\end{equation}

Most of our tests are done for  the single parameter set
$\beta=2.5$ and $\kappa=0.26$. This  set was chosen to compare 
our results for Wilson loops of various sizes with those recently given 
by Irving and Sexton \cite{IrSe}. One simulation was performed at 
$\beta=2.5$ and $\kappa=0.266$ to see the effects of going closer 
to $\kappa_c$.

As elementary updates at level 1 we took 
updates of randomly selected link variables.
These link variables are updated by heat-bath which was
implemented by a multi-hit Metropolis update. 
As criterion to stop the Taylor series of $M^{-1/r} \eta_i$ we used 
\begin{equation}
\label{stop}
\frac{|c_t H^t \eta_i  |^2}{|M^{-1/r} \eta_i |^2}   < 10^{-8} \;\;,
\end{equation}
where $c_t$ is the Taylor coefficient of the order $t$ and 
$M^{-1/r} \eta_i$ is evaluated up to order $t$.

\subsubsection{Testing noise reduction}
In a first set of numerical experiments we studied the effect of 
preconditioning and noise reduction 
on acceptance rates. In order to keep things (conceptually) simple
we used only a two-step decomposition (l=2) for these studies and made no
use of the hopping parameter expansion.

In order to obtain the acceptance rate as a function of the number 
of link variables that are updated in a single proposal we performed 
in a single simulation between 200 and 1000 sequences of update-cycles,
where  in each
sequence the number of updated links for one proposal runs from 1 to 
some maximal number. The acceptance rates are collected during the run.

The simulations are all performed on a $16 \times 16$ lattice. We 
tested the cases of no preconditioning and $r=1$,
and red-black preconditioning
in combination with $r=1,2,4$ and $8$. Our results are given in figure 1.
We made no effort to compute error-bars. The errors should be roughly 
of the size of the fluctuations of the curves.

$50 \%$ acceptance is reached for no preconditioning with $r=1$
at about 2 links updated, while for preconditioning with $r=1$ $50 \%$
acceptance is obtained for about $4$ to $5$ links updated. 
We should note that this amounts to a performance 
advantage of preconditioning by a factor of 4 to 5 since the 
computation of the action requires only  half of the operations that 
are needed in 
the non-preconditioned case.

With preconditioning and $r=2,4,8$ we obtain $50 \%$ acceptance with 
about 9, 20 and 40  links updated. Since the numerical  effort of 
computing the action grows linearly in $r$ there is 
no direct performance gain
by using the roots of the fermion matrix.

However as we will see later it is quite useful that a larger number 
of links can be updated in one update proposal.  The roots might also
allow for a simple version of parallelisation: The application of $H$ 
could be done independently by one processor for each of the $\eta_i$.

We should note of course that the acceptance rate as a function of $r$
is bounded by the acceptance rate that is obtained 
with the exactly evaluated fermion
determinant. Ref. \cite{Langundco1} however suggests that for lattices 
of the size that we consider here even for full sweeps over the lattice
reasonable acceptance rates are obtained  when the  
fermion determinant is evaluated exactly. 
We produced almost independent configurations
by updating 6400 links (which is several times the number of links
of the lattice) in one proposal. We obtained an acceptance rate of 
$0.39(2)$.

\begin{figure}
\begin{center}
\parbox[t]{.85\textwidth}
 {
\caption[acc]
{\label{acc}
Acceptance rates as function of the number of link variables that 
are updated in for the proposal.  For details see the text.
}
}
\end{center}
\end{figure}

\subsubsection{Exploiting the hopping parameter expansion}

In a second set of numerical experiments we studied the inclusion 
of the hopping parameter expansion to order $k=2$ and $k=4$.
Again only a two-step decomposition (l=2) of the action is used.
We performed the test with $r=1$ and $r=4$.  

For $r=1$ on the $16 \times 16$ lattice and preconditioning we find 
$50 \%$ acceptance for $k=4$ at about 30 links updated, for $k=2$ 
at about 13 links updated. For $k=0$ we found already above 4 to 5 links. 
This means that the performance is increased by a factor more than 6 
by using the hopping parameter expansion up to the order $k=4$.

For $r=4$ on the $16 \times 16$ lattice and preconditioning we find 
$50 \%$ acceptance for $k=4$ at about 95 links updated, for $k=2$ 
at about 50 links updated. For $k=0$ we found already above 20 links. 

Hence the performance of the algorithm is more then doubled when $r=2$ 
is used instead of $k=0$ and becomes four-fold for $k=4$.

We did not investigate the inclusion of higher orders of the
hopping parameter
expansion into the algorithm. The order $k=4$ is 
accomplished by a shift of the $\beta$ to $\beta'=\beta+16 \kappa^4$.

Since higher orders of the hopping parameter expansion require new terms
in the gauge-action there will be a trade-off between the evaluation 
of these terms and a larger number of links that can be updated with 
action $S_1$ at a given acceptance rate.

\subsubsection{Using a sequence of actions with $l>2$}
Finally for $r=4$, $k=4$ fixed and red-black preconditioning 
we studied the performance of update
cycles with $l>2$.

The first question that arises is how many levels one should choose and how
one should optimise the parameters $t_i$ (truncation order) and $m_i$
(number of updates at level $i$) of the cycle.

The most direct criterion to judge the quality of an algorithm is 
the product of the square of the statistical error
of the observable that is  
measured multiplied with the CPU-time needed to obtain the  result.
However this criterion requires to perform full simulations  for each
parameter set to be tested.
Therefore one would like to have a more practical
method to estimate the performance of a given cycle.
It seems very natural to assume that the autocorrelation times are 
proportional to the number of links that have been updated
within one cycle.
On the other hand 
the major part of the CPU-time is spent with the 
multiplication
of the off-diagonal part of the fermion matrix $H$ on a vector. 
Both these numbers can be determined with reasonable
accuracy from rather short runs (order 10 cycles).
Therefore we take the ratio of accepted link-updates  divided by the 
number of  $H$ times vector applications as performance index (PI)
of a cycle.

We made a first attempt to perform the optimisation of the 
cycle-parameters automatically.
We start from a guess for the parameters $t_i$ and $m_i$.
The performance index (PI) is computed by averaging over 10 or 20 cycles. 
In a step of the optimisation
we propose small random changes of the $t_i$ and $m_i$ to obtain a 
new proposal  $t_i'$ and $m_i'$. The PI of the $t_i'$ and $m_i'$ is 
computed by averaging over 10 or 20 cycles.
If the PI
of the new parameters is larger than the old one,  $t_i$ and $m_i$ 
are replaced by $t_i'$ and $m_i'$. Typically we performed 200 steps 
in this procedure.
Fortunately it turns out that the performance does not depend very sharply
on the parameters of the cycle. 

\begin{table}[ht]
\caption{\label{schwinger16r4p}
\sl Cycle-parameters for the simulations of the 2D Schwinger 
model at $\beta=2.5$, 
$\kappa=0.26$ on a $16 \times 16$ lattice. For details see the test.}
\vskip 0.2cm
\begin{center}
\begin{tabular}{|c|c|c|c|c|}
\hline
run & level &  $t_i$ & $m_i$ & Accepted \\  
\hline 
1 & 1 &  &   120  &  120   \\
  & 2 & 120.5(7) & 1 &  58.2(5)  \\
\hline 
2 & 1 && 220 & 2640 \\
  & 2 & 14 & 12& 993.6(5.7)  \\
  & 3 & 124.68(39) & 1 &782.7(6.9)  \\
\hline
3 & 1 &  &     120  &  4200 \\
  & 2 &  6  &    5   &   2192.7(5.1) \\
  & 3 &  25  &   7    &   1333.6(6.6) \\
  & 4 & 125.35(35) &  1  &   1249.8(7.3) \\
\hline
4 & 1 &   &    230  & 4600\\
  &  2&  3 &     1   & 2775.2(7.6)\\
  & 3 & 18&     20  & 1318.2(7.3) \\
 & 4& 125.71(40)& 1 & 1119.2(8.6) \\
\hline
\end{tabular}
\end{center}
\end{table}

In a first set of experiments we checked the dependence of the 
performance on the number of cycles. We simulated a $16\times16$
lattice at
$\beta=2.5$ and $\kappa=0.26$.
The truncation orders $t_i$ and the number of applications $m_i$ are 
summarised in table \ref{schwinger16r4p}.  

The $t_i$ refer to the 
expansion of the red-black preconditioned fermion matrix. Hence 
the corresponding order in the hopping parameter expansion is $2 t_i$.
The truncation order of the last level is determined by the truncation 
criterion eq. (\ref{stop}).
In addition in the last column we give
the total number of link-updates per cycle that are accepted at a 
given level. At the first level this is just the total number of
link-updates per cycle. The number given for the last 
level is the number of link-updates that eventually is accepted 
in one cycle.

The statistics of the first run ($l=2$) was 20000
cycles, where the first 2000 are discarded in the analysis.
The statistics of 
the other 3 runs ($l>2$) was 10000 cycles each, and the first 
500 cycles are discarded.

\begin{table}[ht]
\caption{\label{schwinger16r4tau} 
\sl Performance index (PI) and autocorrelation times from simulations
of the 2D Schwinger 
model at $\beta=2.5$ and $\kappa=0.26$
on a $16\times16$ lattice}
\vskip 0.2cm
\begin{center}
\begin{tabular}{|c|c|c|c|c|c|c|c|}
\hline
run & PI   & $\tau_{1\times1}$ & $\tau_{2\times2}$ &
	     $\tau_{3\times3}$ & $\tau_{4\times4}$ &
	     $\tau_{5\times5}$ & $\tau_{stop}$ \\
\hline
1 & 0.0643(6)& 7.6(6)& 9.6(8)&10.5(9)&8.3(7)&5.6(5)& 7.1(5) \\
2 & 0.481(5) & 1.27(7)& 1.41(8)& 1.38(7)& 1.26(8)& 1.01(5) &   \\
3 & 0.523(3)& 0.73(3)&0.81(4)&0.81(4)&0.86(4)&0.72(4) & 0.77(5) \\
4 & 0.576(5)&1.00(5) &1.16(6) &1.14(6)&0.98(5)&0.90(5) &0.92(10)\\
\hline
\end{tabular}
\end{center}
\end{table}

In table \ref{schwinger16r4tau} we summarise the autocorrelation 
times of the Wilson loops of sizes $1\times 1$ up to $5 \times 5$.
In addition we give in the second column the performance index of the run.
And in the last column we give the autocorrelation time of the order 
$t_{stop}$  at which the Taylor series of $M^{-1/r} \eta_i$ is truncated.
This number should be strongly correlated with the small eigenvalues
of $M M^{\dag}$.
Comparing the four runs
we observe that a larger number of accepted link-updates per cycle
indeed corresponds to a smaller autocorrelation time. However comparing
run 1 and 2 we see that the number of accepted link-updates increases 
by a factor of $13.4$ but autocorrelation times only decrease by a factor
of about $7$.

\begin{table}[ht]
\caption{\label{schwinger16r4W} 
\sl  Results for Wilson loops of  size $1\times1$ up to $5\times5$ 
for $\beta=2.5$, $\kappa=0.26$ and $L=16$. The results are summed over
the lattice.}
\vskip 0.2cm
\begin{center}
\begin{tabular}{|c|c|c|c|c|c|c|}
\hline
run   & $W_{1 \times 1}$ &  $W_{2 \times 2}$ &  $W_{3 \times 3}$ &
        $W_{4 \times 4}$ &  $W_{5 \times 5}$ \\
\hline
\cite{IrSe}&201.5(2)  &105.2(6)   &40.5(7)   &12.9(6)   & 3.6(4)   \\
 1  &201.61(14)&105.48(45) &41.20(61) &13.63(50) & 4.34(36) \\
 2  &201.57(8) &104.99(24) &40.61(30) &13.34(27) & 3.97(21) \\
 3  &201.51(6) &105.32(19) &41.08(23) &13.63(22) & 4.19(18) \\
 4  &201.60(7) &105.35(23) &41.29(28) &13.74(24) & 4.29(20) \\
\hline
\end{tabular}
\end{center}
\end{table}

Comparing the performance index we see an improvement by a factor of
$7.5$ from the two-level scheme to the three level scheme.
Going further to four levels one gains about $20\%$ in performance.
Here one should note that the imperfection of the optimisation of
the cycle parameters is a source of uncertainty.

In table \ref{schwinger16r4W} we  give the results for Wilson loops
of the size $1\times1$ up to $5\times5$ summed over the lattice.
In the first line we give for comparison the corresponding
results of ref. \cite{IrSe} which were obtained with hybrid Monte Carlo.
The results are consistent with each other.

\begin{table}[ht]
\caption{\label{schwinger32r4p}
\sl Cycle-parameters for the simulations of the 2D Schwinger
model at $\beta=2.5$,
$\kappa=0.26$ on a $32\times32 $ lattice.}
\vskip 0.2cm
\begin{center}
\begin{tabular}{|c|c|c|c|c|}
\hline
run & level &  $t_i$ & $m_i$ & Accepted \\  
\hline
 1 & 1 &  & 200 & 12000\\
   & 2 & 5 & 1 & 5061.(12.)\\
   & 3 & 20 & 10  & 3081.(10.) \\
   & 4 & 45 & 6 & 2778.(10.)\\
   & 5 & 124.48(19) & 1 & 2749.(11.) \\
 \hline
 2 & 1 &   & 150  & 32400  \\
   & 2 & 7  & 6 & 14369.(21.)  \\
   & 3 & 20 & 6 & 8260.(20.)  \\
   & 4 & 45 & 6 & 7184.(23.) \\ 
   & 5 & 124.28(14) & 1 & 7104.(24.)  \\
\hline
\end{tabular}
\end{center}
\end{table}

Next we tested the dependence of the performance on the lattice size.
Therefore we simulated the $32\times32$ lattice at $\beta=2.5$ and 
$\kappa=0.26$ with $r=4$. We tested two different cycles with $l=5$. 
The parameters of the runs are summarised in table 
\ref{schwinger32r4p}. In table \ref{schwinger32r4tau} we give 
the performance index (PI) and the autocorrelation times of the measured 
observables. 
The performance index is slightly better then 
for $L=16$. However we should note that we still have to take 
into account the $Volume$ dependence of the cost to apply $H$ on a vector.
Therefore we find essentially the expected $Volume^2$ 
dependence of the cost at fixed $\beta$ and $\kappa$ of the algorithm.

\begin{table}[ht]
\caption{\label{schwinger32r4tau}
\sl Performance Index (PI)  and autocorrelation times from 
the simulations at 
$\beta=2.5$ and
$\kappa=0.26$ on a $32\times32$ lattice.}
\vskip 0.2cm
\begin{center}
\begin{tabular}{|c|c|c|c|c|c|c|c|}
\hline
run & PI   & $\tau_{1\times1}$ & $\tau_{2\times2}$ &
	     $\tau_{3\times3}$ & $\tau_{4\times4}$ &
	     $\tau_{5\times5}$ & $\tau_{stop}$ \\
\hline
 1 &0.604(3) &0.89(6)& 1.05(7)& 1.13(8)& 1.02(6)& 0.69(5)& 0.97(6) \\
 2 &0.703(3) & 0.60(2)& 0.63(3)&0.63(3)&0.58(3) &0.52(2)&0.59(2) \\
\hline
\end{tabular}
\end{center}
\end{table}

The results for Wilson loops of size $1\times1$ up to $5\times5$
are summarised in
table \ref{schwinger32r4W}. The results of our runs are consistent. 
However there is some mismatch with the data of ref. \cite{IrSe}. 
In particular the value for the $2\times2 $ Wilson loop is by 4.7 
standard deviations smaller than the combined result from our simulations.
Note that our result for the $32\times32$ lattice is consistent with 
the results obtained for the $16\times16$ lattice.

\begin{table}[ht]
\caption{\label{schwinger32r4W}
\sl
 Results for Wilson loops of  size $1\times1$ up to $5\times5$ at
 $\beta=2.5$,
$\kappa=0.26$ on a $32\times32$ lattice.}
\vskip 0.2cm
\begin{center}
\begin{tabular}{|c|c|c|c|c|c|c|}
\hline
run   & $W_{1 \times 1}$ &  $W_{2 \times 2}$ &  $W_{3 \times 3}$ &
        $W_{4 \times 4}$ &  $W_{5 \times 5}$ \\
\hline
 \cite{IrSe}&805.1(3)   &415.7(9)  & 158.9(11)  &51.2(10) & 15.1(10)  \\
1 &805.79(16) &419.62(52) &162.12(68) &52.79(55) &16.12(40) \\
2 &805.95(14) &420.52(42) &163.83(54) &54.44(48) &17.11(35) \\
\hline
\end{tabular}
\end{center}
\end{table}
\begin{table}[ht]
\caption{\label{schwinger32px}
\sl Cycle-parameters for the simulations of the 2D Schwinger
model at $\beta=2.5$ and
$\kappa=0.266$ on a $32\times32 $ lattice. }
\vskip 0.2cm
\begin{center}
\begin{tabular}{|c|c|c|c|}
\hline
 level &  $t_i$ & $m_i$ & Accepted \\  
\hline
 1 &    &150 & 32400 \\
 2 &  6 &1  & 11654.(23.) \\
 3 & 16 &6 & 7355.(19.) \\
 4 & 40 &6 & 5073.(16.) \\
 5 & 80 &6 & 4377.(19.) \\
 6 & 283.5(1.5)  &1  & 4270.(22.) \\
\hline
\end{tabular}
\end{center}
\end{table}

Finally we performed one run on a $32 \times 32$ lattice 
at $\beta=2.5$ and  $\kappa=0.266$  in order to check the dependence 
of the performance on $\kappa$. Note that for $\beta=2.5$ 
$\kappa_c \approx 0.272$ from interpolating the results given in table 
1 of ref. \cite{Langundco2}.
The simulation consists of 9000 cycles.
The first 500 cycles were discarded
from the data-analysis. The simulation (which was the most expensive 
discussed in this section) took about 7 days of CPU on a 200 MHz Pentium
Pro PC. The parameters of the cycle are given in table 
\ref{schwinger32px}. We see that the number of terms to compute 
$M^{-1/r} \eta_i$ increases by a factor of $2.28$ compared with 
$\kappa=0.26$.

\begin{table}[ht]
\caption{\label{schwinger32taux}
\sl Performance Index (PI)  and autocorrelation times from 
the simulation at 
$\beta=2.5$ and
$\kappa=0.266$ on a $32\times32$ lattice.}
\vskip 0.2cm
\begin{center}
\begin{tabular}{|c|c|c|c|c|c|c|}
\hline
PI   & $\tau_{1\times1}$ & $\tau_{2\times2}$ &
	     $\tau_{3\times3}$ & $\tau_{4\times4}$ &
	     $\tau_{5\times5}$ & $\tau_{stop}$ \\
\hline
0.254(13) &0.80(4)&0.85(5)&0.86(5)&0.80(5)&0.59(3)& 0.90(7)\\
\hline
\end{tabular}
\end{center}
\end{table}
In table \ref{schwinger32taux} we give the autocorrelation times
of Wilson loops and the truncation order of the Taylor series 
at the final level. In the first column we give the performance 
index. The performance index degrades by a factor of $2.77$ compared
with our best cycle for $\kappa=0.26$ and $L=32$. This means that 
in addition to the larger costs for evaluating  $M^{-1/r} \eta_i$
there is a small degradation of the performance due to 
reduced acceptance rates.

\begin{table}[ht]
\caption{\label{schwinger32Wx}
\sl Results for Wilson loops of size $1\times1$ up to $5\times5$ at 
$\beta=2.5$ and
$\kappa=0.266$ on a $32\times32$ lattice. The results are summed
over the lattice.}
\vskip 0.2cm
\begin{center}
\begin{tabular}{|c|c|c|c|c|c|c|}
\hline
 $W_{1 \times 1}$ &  $W_{2 \times 2}$ &  $W_{3 \times 3}$ &
 $W_{4 \times 4}$ &  $W_{5 \times 5}$ \\
\hline
 810.21(14) &435.28(40) &182.76(51) &68.59(47) &25.09(35) \\
\hline
\end{tabular}
\end{center}
\end{table}

In table \ref{schwinger32Wx} we give the results for 
Wilson loops of size $1\times1$ up to $5 \times 5$. 
In particular  the 
values of the large loops are considerably larger than for 
$\kappa=0.26$.

\subsection{2 flavour QCD with Schr\"odinger functional boundary
	    conditions}
We simulated the standard Wilson gauge action with two flavours of 
mass-degenerate Wilson fermions.
We performed runs at $\beta=8.3$ and $\kappa=0.1386 \approx \kappa_c$
on a
$8^4$ lattice. We applied Schr\"odinger functional boundary 
conditions \cite{schrodinger1,schrodinger2,sint,schrodinger3}.
The gauge-fields at the boundaries are chosen as specified 
in ref. \cite{schrodinger2} with $c_t=1.0$. 
The boundary conditions for the fermions 
are taken as specified in ref. \cite{schrodinger3} with 
$\theta=\pi/5$. 

The Schr\"odinger functional boundary conditions and the particular 
set of parameters were chosen in order to compare the performance of the 
algorithm with that of the 
hybrid Monte Carlo and the non-hermitian version of the multi-boson
algorithm 
which are benchmarked by the Alpha collaboration at these
parameters. 

As observables we have implemented the plaquette, the inverse of the 
running coupling $\bar{g}^{-2}$ and $\bar{v}$ 
(for the definitions of these
quantities see ref. \cite{schrodinger2}).  In addition we computed the 
autocorrelation time of the number of iterations needed for convergence 
of the Taylor series. This number should be closely related to the 
smallest eigenvalues of $M M^{\dag}$. 

Biased by the results of the previous section 
we used red-black preconditioning and 
the hopping parameter expansion to order
$k=4$ for our simulations. 
To this order the hopping parameter expansion leads to a shift
in $\beta$
\begin{equation}
 \Delta \beta=96 \;\; \kappa^4  \;\; .
\end{equation}

For our QCD simulations we tried to do better than randomly selecting 
the links to be updated at level 1 of the update cycle.
We sweep through  sub-blocks of a certain size in an ordered way.
Here one should note that for fixed auxiliary bosonic fields $\eta$ 
the action
$S_l$ is not gauge invariant. Therefore fixing the gauge should increase
acceptance rates.

As elementary link-updates we used
Cabibbo-Marinari heat-bath updating and micro-canonical over-relaxation.
In the case of Cabibbo-Marinari we performed 
a sequence of 5 $SU(2)$-subgroup  heat-bath updates 
where the
subgroups are given by the $(1,2)$, $(2,3)$, $(1,3)$, $(2,3)$ , $(1,2)$
components of the $SU(3)$ matrix. Also in the case of over-relaxation 
we updated in a sequence of $SU(2)$ subgroups. The sequence
is given by
the $(1,2)$, $(2,3)$, $(1,3)$ components of the $SU(3)$ matrix. 
While in the case of Cabibbo-Marinari our sequence is symmetric, 
symmetrisation as discussed in section 3.3 has to be done  
in the case of over-relaxation.

We tested two different update schemes for the update of the action
$S_1$.

In our first update-scheme,
which will in the following be refered to as "point-update", we choose
with uniform probability a lattice point with $0<t<T$. Then  7
of the links attached to that point 
are updated as explained below. The $8^{th}$ link is
kept fixed in order to avoid updating of gauge degrees of freedom.
We sweep 5 times over the 7 links, where the single link variable is 
updated by a Cabibbo-Marinari heat-bath update. The order of sweeping 
through the 7 links is again symmetrised.

Since the links in time direction at the boundaries $t=0$ and $t=T$ do
not couple to the fermions, updates of these links come almost for free.
Therefore we performed a lexicographic sweep with over-relaxation 
updating over all links in time direction at 
the boundaries after half of the point-updates are performed.
Also here the sequence of the updating is symmetrised. Note that most 
of the specifications of the details of the update are taken ad hoc.

The second scheme that we tested,  
which will in the following be refered to as "block-update" is 
characterised as follows.
First a  sub-block of size $4^4$ is selected. The position in the 
spatial directions is chosen randomly with an uniform distribution. 
In temporal direction the block is 
either attached to the $t=0$ or the $t=T$ boundary; i.e. the block runs 
either from $t=1$ to $t=4$ or from $t=4$ to $t=7$ (with equal probability
for the two cases).  

In order to avoid updating gauge degrees of freedom,  
only spatial links and temporal links at the boundaries $t=0$ and $t=T$
are updated. 

The update-sequence for a given sub-block is the following:
First a sweep in lexicographic order
with Cabibbo-Marinari through the spatial
links of the sub-block is performed.
Then there are 8 over-relaxation sweeps over the spatial links of the 
sub-block and the temporal links of the sub-block at the boundary.
Finally there is a heat-bath sweep over all temporal boundary links.
With  probability one half the exact reverse 
of this sequence is performed in order to
satisfy detailed balance.

In addition to the runs with dynamical fermions we performed simulations
with the pure gauge action using the two updating schemes discussed
above. The idea is that the Monte Carlo dynamics of the quenched 
simulation is very similar to that of the simulation of the full theory.

Then we could obtain an estimate of the performance of the full algorithm
by combining autocorrelation times of the quenched simulation 
with acceptance rates of the dynamical fermion simulation. 
Acceptance rates can be 
obtained reasonably well from rather short runs, while reliable estimates
of auto-correlation times require rather good statistics.
\begin{equation}
\tau_{fermions} \approx \frac{n_{acc}}{n_{quenched}} \tau_{quenched} \;\;, 
\end{equation}
where $n_{acc}$ is the number of point-updates or block-updates 
accepted in an update cycle of the dynamical fermion simulation and 
$n_{quenched}$ is number of point-updates or block-updates which is 
performed per measurement in the quenched simulation.

\subsubsection{Quenched simulations}
First we performed quenched simulations with the two updating schemes 
discussed above.

For the point-update we performed  20000 measurements. Per measurement
$12 \times 360$ point-updates are performed.
For the block-update we performed 30000 measurements.  
Per measurement we updated 16 $4^4$ sub-blocks. 

In both cases the first 1000
measurements were discarded for the evaluation of the observables
and integrated autocorrelation times which are summarised in table 
\ref{quenchedruns}.

\begin{table}[ht]
\caption{
\label{quenchedruns}
\sl Results for the average plaquette (plaq), the running coupling
$\bar{g}^{-2}$, $\bar{v}$ and the corresponding autocorrelation times 
of the quenched runs. For details see the text.}
\vskip 0.2cm
\begin{center}
\begin{tabular}{|c|c|c|c|c|c|c|}
\hline
update & plaq  & $\tau_p$ & $\bar{g}^{-2}$ & $\tau_g$ & 
$\bar{v}$ & $\tau_v$ \\
\hline
point &0.73963(1)& 0.71(3) & 0.6814(51) & 2.62(13) & 0.075(11)& 8.2(9) \\
block &0.73963(1)& 1.36(4) & 0.6742(31) & 1.52(4)  & 0.062(7) & 3.8(2) \\
\hline
\end{tabular}
\end{center}
\end{table}
We observe that the integrated autocorrelation time of $\bar{v}$ is the 
largest of the measured autocorrelation times. 

\subsubsection{Simulations with dynamical fermions}
First we tested 
the "point-updating" scheme. 
After some preliminary testing we decided to use $l=6$ for the 
approximation sequence, the root $r=6$ and the order of the
hopping parameter expansion $k=4$.  
The truncation order and the multipicities 
for each level is given in table \ref{run1}.
In table \ref{run1} we give in addition 
the average total number of point-updates that are accepted at
a given level during an update-cycle. At level 1 this is just the 
total number of "point updates" in one cycle
\begin{equation}
21600=360 \times 1 \times 1 \times 10 \times 6 \times 1 \;\;.
\end{equation}
In order to obtain a reasonably large statistics we used trivial 
parallelisation.
We performed 6 independent runs, 
where the first 400 cycles of each run are discarded for the data
analysis. 
All runs were started from 
an ordered configuration. Then a small number of update-sweeps with 
the pure gauge action plus the $\Delta \beta$ shift was performed in order 
to avoid 
convergence problems of the Taylor series.
In total we generated 8450 measurements that were used for the averaging.
The  simulation
took about 135 days of Pentium Pro 200 MHz time, where
the generation of the discarded configurations is not taken into account.

\begin{table}[ht]
\caption{\label{run1} \sl Cycle-parameters of the dynamical fermion simulation
 with the "point-update" scheme.}
\vskip 0.2cm
\begin{center}
\begin{tabular}{|c|c|c|c|}
\hline
level &  $t_i$  &   $m_i$  & accept \\
\hline
 1    &  Gauge + hopping   &        360      &      21600       \\
 2    &         4          &          1      &       8844(27)   \\
 3    &	        7          &          1      &       6178(25)   \\
 4    &        10          &         10      &       5386(24)   \\
 5    &        24          &          6      &       4675(24)   \\
 6    &    "$\infty$"      &          1      &       4548(24)   \\
\hline
\end{tabular}
\end{center}
\end{table}
The CPU-costs of an update-cycle are essentially given by the number 
of applications of the off-diagonal part of the the fermion matrix
to a vector. For our run 
on average 4075(3) applications of $H$ per cycle were performed.

The 
average  truncation order of the Taylor series is 98.5(1). 
The autocorrelation time of the (over the 6 replicas averaged)
truncation order is $\tau_{stop} =  2.4(2)$. Results for the observables
are given in table \ref{obs1}.

\begin{table}[ht]
\caption{\label{obs1} 
\sl Results for the average plaquette (plaq), the running coupling
$\bar{g}^{-2}$, $\bar{v}$ and the corresponding autocorrelation times 
of the dynamical fermion simulation with the "point-update" scheme.
For details see the text.}
\vskip 0.2cm
\begin{center}
\begin{tabular}{|c|c|c|c|c|c|}
\hline
  plaq  & $\tau_p$ &  $\bar{g}^{-2}$ & $\tau_g$ &  $\bar{v}$ & $\tau_v$ \\
\hline
 0.742746(18) &  0.83(4) &  0.7157(73) & 2.47(20) & 0.096(16) &6.9(9)\\
\hline
\end{tabular}
\end{center}
\end{table}

First we note that the results for the observables are consistent 
with those obtained by U.Wolff with the multi-boson algorithm 
\cite{uprivate}. As for the quenched simulation
the autocorrelation time of the plaquette is the smallest
while that of $\bar{v}$ is the largest. It is even considerable larger 
than  $\tau_{stop}$. Next we checked our hypothesis that the 
autocorrelation times can be predicted from autocorrelation times 
of the quenched run in combination with acceptance rates of the dynamical 
fermion simulation.
The numbers of table \ref{quenchedruns} have to be multiplied 
with $4320/4548$. Then we obtain the estimates $\tau_{p,est} =0.67(3)$,
$\tau_{g,est} =2.49(12)$  and $\tau_{v,est} = 7.8(9)$
which is in quite good 
agreement with the directly measured results given in table \ref{obs1}.

The Alpha collaboration  intents to compute the running coupling of 
QCD with dynamical fermions.
Therefore the comparison of the algorithms 
which are candidates for this project  is based on the 
statistical error of $\bar{g}^2$.
We agreed on the following cost definition
\begin{equation}
D_{cost} := (\mbox{total number of applications of } H)  \times
	     \mbox{Error}[\bar{g}^{-2}]^2
\end{equation}

For the run with the "point-update" we obtain 
\begin{equation}
D_{cost} = 
1835(150)+181(15) \;\; ,
\end{equation}
where the error of $D_{cost}$ is derived from the statistical
error of $\tau_g$. The
second contribution comes from converting the updating costs at level $1$. 
The cost of the Cabibbo-Marinari updates at level 1 have been converted to units of $H$
application and are given by the second number.

In a similar way as for the autocorrelation times we tried to 
estimate the $D_{cost}$ form the statistical error of $\bar{g}^{-2}$ 
in the quenched simulation and the total acceptance in the 
dynamical fermion simulation. We obtain
\begin{equation}
 D_{cost,est}=1913(95)+189(9) \;\; ,
\end{equation}
which is again in good agreement with the directly obtained result.

Next we studied the "block-update" scheme. Here we also used $l=6$,$r=6$ 
and $k=4$. The truncation orders and the multiplicities are 
summarised in table \ref{block}. We performed only a run of 100 cycles
that was started from an equilibrium configuration. 
The total acceptances are also given in table  \ref{block}.
\begin{table}[ht]
\caption{\label{block} \sl Cycle-parameters of the dynamical fermion simulation
 with the "block-update" scheme.}
\vskip 0.2cm
\begin{center}
\begin{tabular}{|c|c|c|c|}
\hline
level &  truncation order  &   multiplicity  & total acceptance \\
\hline
 1    &  Gauge + hopping   &          1      &      60         \\
 2    &         4          &          1      &      34.3(5)    \\
 3    &         8          &          3      &      25.7(4)    \\
 4    &        14          &          4      &      22.2(4)    \\
 5    &        25          &          5      &      20.1(5)    \\
 6    &    "$\infty$"      &          1      &      19.9(6)    \\
\hline
\end{tabular}
\end{center}
\end{table}

On average a cycle of the update requires 4836(16) applications  
of the off-diagonal part of the fermion matrix.
The 100 measurements are not sufficient to produce reliable estimates of
the autocorrelation times and of the statistical error of $\bar{g}^{-2}$
Therefore we only quote the estimated $D_{cost}$:
\begin{equation}
D_{cost,est}=1084(50) + 62(3) \;\; .
\end{equation}
We conclude that "block-update" is more efficient than the "point-update".

Our $D_{cost}$ results can be compared with  the other two 
algorithms benchmarked by the Alpha collaboration:
\begin{equation}
D_{cost} \approx 900(30) 
\end{equation}
was obtained as preliminary result by U.Wolff with the non-hermitian 
version of the multi-boson algorithm with re-weighting  
\cite{uprivate} and
\begin{equation}
D_{cost} \approx 460
\end{equation}
was obtained by K. Jansen \cite{kprivate} with the polynomial hybrid
Monte Carlo algorithm \cite{phmc}. In the last case only $H$ applications
have been counted. Additional costs have not been converted into 
$D_{cost}$.

We conclude that for a $8^4$ system we already have reached similar
efficiency as the multi-boson and the polynomial 
hybrid Monte Carlo algorithm. However, at least with the present 
implementation, we have to expect a worse scaling  
of the 
efficiency with the lattice size 
as for the multi-boson and the hybrid Monte Carlo algorithm.
Therefore further progress is needed to obtain a competitive 
algorithm for
lattice sizes  that are needed for the calculation for the 
running coupling 
or light hardron spectroscopy.

\section{Comparison with related approaches}
Irving and Sexton \cite{IrSe} 
use approximations of the fermion determinant 
to simulate the 2D Schwinger model and QCD. 
They try to approximate the fermion action by adding 
Wilson-loops of different sizes and shapes to the gauge action.
Among other things
they propose to construct an exact algorithm by generating 
a proposal by using the approximate action.
They have in mind a two-level Metropolis scheme. However also in 
their context one could think of a sequence of approximations by 
incorporating more and more Wilson loops.

In a recent paper Duncan, Eichten and Thacker \cite{DuEiTh} 
propose to split the fermion determinant into two parts. 
One part is given by the product of the smallest eigenvalues. 
They suggest that the remaining part can be expressed in terms of 
small Wilson loops. In their simulation they 
approximated the fermion determinant just by the product of the 
the smallest eigenvalues. They argue that the remaining part of 
the the fermion determinant effectively amounts to a $\beta$-shift.

In the same spirit one could simulate with a $\tilde M$ 
that corresponds
to a moderate order of the hopping-parameter expansion and  
ignor  the small orders of the hopping parameter expansion.

Thron, Dong,  Liu and Ying  \cite{pade} estimate the 
fermion determinant stochastically. Their method to reduce the noise
of their estimator incorporates, as $\tilde M$ in this 
paper, the hopping parameter expansion.  


\section{Outlook}
There are several directions in which progress could be made.

First one could try to use
higher orders of the hopping-parameter expansion.
In the present paper we only considered the hopping parameter 
expansion up to order $k=4$. This leads only to a shift in $\beta$.
At order $k=6$, in 
4 dimensions, Wilson loops of three different shapes contribute. 
These were implemented for example in ref. \cite{pade}. Going to even 
higher orders, the number of Wilson loops needed grows exponentially in 
the order. Therefore one has to look for a more efficient method 
to compute 
$\mbox{tr} H^k $ than expressing it in terms of Wilson loops. 


One might find noise reduced unbiased estimators of the
fermion determinant that are less 
CPU-intensive than  those discussed in section 2.2 of this paper.

Finally one might 
look at approximation schemes different from that discussed 
in this paper. In this paper the Taylor series of the fermion 
matrix in $\kappa$ is used as the basis of the approximation scheme.
One might construct approximations of the action 
by truncating $M^{-1}$ in real space,
i.e. allowing only non-zero matrix elements $(M^{-1})_{xy}$ with 
$|x-y|< d$. The quality of the approximation is then controlled by $d$.

\section{Acknowledgements}
I would like to thank  the Alpha-collaboration for providing
Fortran90-code for pure QCD with Schr\"odinger functional 
boundary conditions
(Stefan Sint, Stefano Capitani), performance data  for the multi-boson
and hybrid Monte Carlo Algorithm (Ulli Wolff, Karl Jansen) 
and finally 
for discussions with all participants of the tuesday seminar in our 
group at Humboldt University, Klaus Pinn and Stefano Vinti.

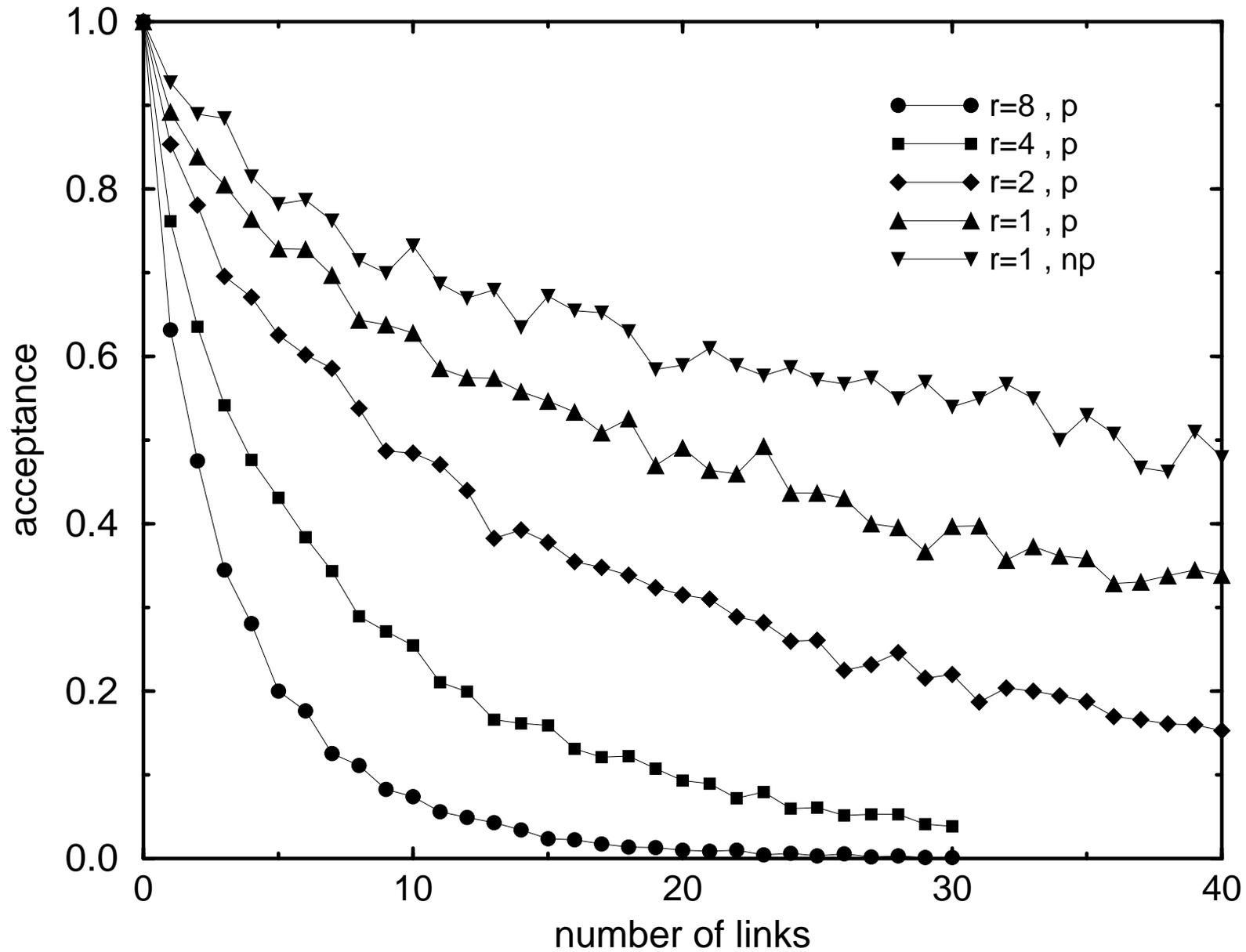
\end{document}